# Uncertain Congestion Games with Assorted Human Agent Populations

**Asrar Ahmed, Pradeep Varakantham, Shih-Fen Cheng**
School of Information Systems, Singapore Management University, Singapore 178902
{masrara,pradeepv,sfcheng}@smu.edu.sg

## Abstract

Congestion games model a wide variety of real-world resource congestion problems, such as selfish network routing, traffic route guidance in congested areas, taxi fleet optimization and crowd movement in busy areas. However, existing research in congestion games assumes: (a) deterministic movement of agents between resources; and (b) perfect rationality (i.e. maximizing their own expected value) of all agents. Such assumptions are not reasonable in dynamic domains where decision support has to be provided to humans. For instance, in optimizing the performance of a taxi fleet serving a city, movement of taxis can be involuntary or non-deterministic (decided by the specific customer who hires the taxi) and more importantly, taxi drivers may not follow advice provided by the decision support system (due to bounded rationality of humans).

To that end, we contribute: (a) a general framework for representing congestion games under uncertainty for populations with assorted notions of rationality. (b) a scalable approach for solving the decision problem for perfectly rational agents which are in the mix with boundedly rational agents; and (c) a detailed evaluation on a synthetic and real-world data set to illustrate the usefulness of our new approach with respect to key social welfare metrics in the context of an assorted human-agent population.

An interesting result from our experiments on a real-world taxi fleet optimization problem is that it is better (in terms of revenue and operational efficiency) for taxi drivers to follow perfectly rational strategies irrespective of the percentage of drivers not following the advice.

## 1 Introduction

Congestion games [5] and its variants have numerous applications in domains ranging from urban transportation [14] (e.g. movement of vehicles between different regions of an area) to capturing company dynamics in a congested industry [12] (e.g., strategizing on marketing investments by different companies selling the same product). There are many interesting challenges that exist in these problem domains: (a) Representing and accounting for implicit interactions between agents. For example, vehicles trying to get on the same road are implicitly competing (even though there may not be an actual intention to compete); (b) large-scale nature of problems; (c) involuntary movements of agents (in some decision epochs, agents might need to *follow* given instructions and cannot freely choose their own actions); and (d) accounting for humans in the loop, who may not follow perfectly rational policies.

While existing research in congestion games is extensive, we are unaware of research that addresses the issues mentioned in points (b), (c) and (d) mentioned above. To that end, we make three key contributions in this paper: (1) By extending on the cognitive hierarchy model [1] (introduced in behavioral game theory), we introduce a general model to represent the decision problem for perfectly rational agents in an agent population of assorted rationalities; (2) A scalable approach called Soft-max based Flow update for Assorted Populations (SoFA) for solving the decision problem of perfectly rational agents in an assorted population; and (3) Finally, we provide a detailed evaluation on both real-world and synthetic data sets for the taxi fleet optimization problem.

This paper is motivated by a real-world problem of improving the performance of a taxi fleet. The decision making problem faced by a typical taxi driver is interesting both practically and theoretically, since a driver needs to make both voluntary (driver's own decision) and involuntary (when customers board taxis) movements. In such a problem, customers are considered as the resources, due to whom an implicit competition exists between taxis. Since demands are both zone-dependent and time-dependent, the problem becomes even more challenging. The goal here is to provide decision support to taxi drivers such that the operational efficiency of the fleet and the revenues obtained are improved. Similar problems exist in analyzing industry dynamics (where different companies strategize to maintain their competitive advantage) and labor mobility (where individuals reason about their movement to different geographical regions). Although the concept of *user equilibrium* [10] is well-adopted in modeling either static or dynamic traffic route selection, it cannot be applied in this class of problems due to the presence of involuntary movements.

Experimentally, we are able to show that SoFA converges on all our problems (both real-world and synthetic ones) with assorted human-agent populations (agents with different levels of rationality). It is an important result since we are able to show that if our algorithm converges, the solution for perfectly rational agents is an equilibrium strategy (i.e., no incentive w.r.t expected value). Another nice empirical result of our approach in the taxi fleet optimization problem is that perfectly rational drivers (i.e. ones who adopt policies suggested by SoFA) always outperform agents who do not follow the suggested policies.

## 2 Motivation: Taxi Fleet Optimization

Our research is motivated by the problem of optimizing a fleet of self-interested taxis[1]. For a fleet of taxi population $\mathcal{P}$, serving a city divided into a set of zones $M$, the goal is to provide decision support for taxi drivers to independently decide a sequence of zones to roam in.

The demand fulfillment is modeled at zone-level, assuming demands are common knowledge among all taxis. If the number of taxis in a particular zone during a time period is fewer than the number of customers, all taxis will be hired (the determination of their destinations is described in the following paragraph). Otherwise, only a fraction of the roaming taxis (up to the number of customers) will be hired.

The movements of taxis between zones depend on whether they are hired or not. If a taxi is hired, the movement is involuntary (decided by the customer on-board) and is governed by the probability distribution computed from the outgoing flows of customers to different zones. Therefore, if a taxi is hired by a customer in zone $i$, the probability of moving to zone $j$ is computed based on the fraction of customers moving from zone $i$ to $j$ over all flows out of zone $i$. Furthermore, a hired taxi in such a case will receive a revenue of $Re^t(i,j)$ and incur a cost of $Co^t(i,j)$. On the other hand, if the taxi is not hired, the movement will be voluntary and deterministic, and the taxi will receive no revenue but incur the same cost of $Co^t(i,j)$.

---
[1] In our definition, we assume that each taxi is driven by an independent driver (thus we use *taxi* and *taxi driver* interchangeably), who pays for all costs (e.g., monthly rent, fuel, maintenance) and keeps all earned revenue. We also assume that there will be no communication/collaboration among taxi drivers.

The goal is to provide decision support (in terms of policies) for drivers who request for decision support on moving through the island, such that there is no incentive for them to deviate from their policies.

## 3 Model: DAAP

We now introduce a general framework for modeling problems such as taxi fleet optimization called the Distributed decision model for Assorted Agent Populations (DAAP). DAAP is an extension of the DDAP (Distributed Decision model for Agent Populations) model introduced by Varakantham *et al.* [7]. DAAP can be viewed as a framework for representing generalized congestion games with movement uncertainty. In particular, we generalize the notion of resources in congestion games to states and the movement uncertainty to transition functions (akin to the one in Markov Decision Problems, MDPs).

DAAP represents a subset of problems represented by the generic stochastic game model [2]. In DAAP, the transition and reward functions for an agent are dependent only on the aggregate distributions of other agent states, whereas in more general models like stochastic games the transition and reward function for an agent are dependent on specific state and action of every other agent. It should be noted that DAAP represents problems with selfish agents and hence is different to cooperative models such as DEC-POMDP.

DAAP is the tuple:

$$\langle \Gamma, \{\mathcal{P}_\tau\}_{\tau \in \Gamma}, \mathcal{S}, \mathcal{A}, \{\phi_\tau\}_{\tau \in \Gamma}, \{\mathcal{R}_\tau\}_{\tau \in \Gamma}, d^0, \{\mathcal{V}_\tau\}_{\tau \in \Gamma} \rangle,$$

where $\Gamma$ represents the different agent types and $\mathcal{P}_\tau$ is the set of agents of type $\tau$. Two agents belong to different types if they have (a) different transition ($\phi$) or reward model ($\mathcal{R}$); or (b) different notion of rationality ($\mathcal{V}$). $\mathcal{S}$ corresponds to the set of states encountered by each agent. $\mathcal{A}$ is the set of actions executed by each agent. We define the set of state distributions, $D = \{d | d = \langle d_1, d_2, \cdots, d_{|\mathcal{S}|} \rangle, \sum_{s \in \mathcal{S}} d_s = |\mathcal{P}|\}$, where $\mathcal{P}$ is the set of all agents and $d_s$ represents the number of agents in state $s$. $d^0$ represents the starting distribution of agent states. $\phi_\tau$ models the involuntary movements of every agent of type $\tau$ and more specifically, $\phi_{\tau,d}^t(s,a,s')$ represents the probability that an agent of type $\tau$ in state $s(\in \mathcal{S})$ after taking action $a(\in \mathcal{A})$ would transition to state $s'$, when the distribution is $d$ and time is $t$. $\mathcal{R}^t_{\tau,d}(s, a, s')$ is the reward obtained by an agent of type $\tau$ when in state $s$, taking action $a$ and moving to state $s'$ when the distribution of agents is $d$ at time $t$.

$\mathcal{V}_\tau$ characterizes the rationality concept of interest to agents of type $\tau$ and represents the value function definition for agents of type $\tau$. This is provided to capture the notion of bounded rationality (particularly for humans in the population). In this paper, we refer to the two extreme cases for bounded rationality, which are:
(a) Perfect rationality: agents compute exact best response to opponent policies;
(b) Local rationality: agents assume no other agent is present ($d^t = \mathbf{0}, \forall t$) in the system.
Further details on various rationality concepts are provided in Section 4.

The objective in solving a DAAP is to compute a policy, $\pi_i$ for each agent $i$ of type $\tau$, such that there is no incentive for any agent to deviate from its policy, in terms of the value (i.e. $\mathcal{V}_\tau(\pi_i, \{\pi_k\}_{k \in \mathcal{P}_\tau, k \neq i})$).

### 3.1 Taxi Fleet Optimization as DAAP

The number of types in this problem is primarily due to varying degrees of human rationality. We will describe this in detail in Section 7.1. In representing the taxi fleet optimization as a DAAP, the key is using the transition function to represent the involuntary movement of taxis[2]. $\mathcal{P}$ is the set of taxis in the fleet, $\mathcal{S}$ is all the zones in which a taxi could be present. $\mathcal{A}$ is the set of zones to which a taxi driver wishes to move. The transition function, $\phi$ is computed based on the customer flow, $fl$ between various zones. $D^0$ is the distribution of taxis at the starting time. Equation 1 provides the expression for computing the transition probability between states.

Intuitively, if there are fewer taxis than customers in a zone, then all taxis are hired and their transition probability to a specific zone is dependent on flows to different zones from the current zone (represented by condition **C1**). If there are more taxis in a zone than customers, then transition is dependent on whether the action (intended zone) is same as the destination zone (captured by conditions **C2** and **C3**).
**C1**: if $\sum_{\hat{s}} fl^t(s, \hat{s}) \geq d_s$
**C2**: if $a \neq s'$, $\sum_{\hat{s}} fl^t(s, \hat{s}) < d_s$
**C3**: if $a = s'$, $\sum_{\hat{s}} fl^t(s, \hat{s}) < d_s$

$$\phi^t_d(s, a, s') = \begin{cases} \frac{fl^t(s, s')}{\sum_{\hat{s}} fl^t(s, \hat{s})} & \textbf{C1} \\ \frac{fl^t(s, s')}{d_s} & \textbf{C2} \\ 1 - \frac{\sum_{\hat{s} \neq s'} fl^t(s, \hat{s})}{d_s} & \textbf{C3} \end{cases} \quad (1)$$

---

[2]Since the movement of taxis and their revenues (standardized meters) are identical, we do not index transition and reward model of DAAP with type.

Similar to the transition, $\mathcal{R}$ is the reward obtained based on the three conditions.

In solving the taxi optimization DAAP, our goal is to maximize expected revenue for the individual taxi drivers while reducing starvation. Since the transition function depends on the number of taxis in the zone, maximizing expected revenue implies minimizing starvation as well. In this problem domain, both the welfare metrics (revenue and starvation) are optimized at once, however, in other domains there could be multiple objectives that are not in alignment and multi-objective reasoning might be required.

## 4 Characterizing Rationality

A major contribution of the paper is the introduction of multiple levels of rationality to the uncertain congestion game that mimics bounded rationality of human decision makers. In the context of DAAP, this implies the computed policy will be followed differently: perfectly rational agents adopt the policy completely, while other agents only adopt the policy partially (or even not at all). To concretely represent bounded rationality, we extend on the cognitive hierarchy model developed by Camerer *et al.* [1]. Therefore, in DAAP, $\mathcal{V}_\tau$ will represent the rationality notion of an agent in terms of the cognitive hierarchy model.

Cognitive hierarchy model assumes there is a bound on the levels of reasoning agents can do and that agents differ in their levels of reasoning. *Level-0* agents are assumed to be non-strategic and either play randomly or use a heuristic. *Level-1* agents compute best response strategy by assuming all other agents in the population are *level-0*. In general, *level-L* agents compute best response strategy by assuming a distribution $f$ across levels $\{0, 1, \ldots, L-1\}$. Camerer *et al* present a single-parameter model, called *Poisson cognitive hierarchy* model, where $f$ is assumed to follow Poisson distribution.

We extend the cognitive hierarchy model in three ways. Firstly, the opponent distribution $f$ is obtained from iterative Bayesian inference (details provided in Section 7.1). Secondly, the best response is computed using the *quantal best response* strategy. The quantal best response for player $i$ to strategy $s_{-i}$ is given by:

$$s_i(a_i) = \frac{e^{\lambda \cdot u_i(a_i, s_{-i})}}{\sum_{a'_i} e^{\lambda \cdot u_i(a'_i, s_{-i})}}.$$

For $\lambda = 0$, each action is assigned equal probability (which can be thought as completely irrational); as $\lambda$ increases, quantal best response approaches standard best response (which is completely rational). Finally, we introduce the consideration of different decision horizons (referred to as $T$). With the above extensions, an agent's behavior can be quantified by a three-tuple CH_QR($\lambda, L, T$). To illustrate the flexibility of our model, we provide two examples below::

**Example 1** *CH_QR(1, 0, 1) corresponds to an algorithm that generates a quantal best response strategy with $\lambda = 1$ that maximizes one-step payoff without considering other agents. CH_QR($\infty, \infty, H$) corresponds to an algorithm that generates the exact best response considering full horizon, assuming all levels of reasoning from other agents.*

This parametric model enables us to evaluate the SoFA technique on various sets of assorted populations and demonstrate the utility of having perfectly rational agents within a human population. The Poisson cognitive hierarchy model has been shown to provide a good fit for empirical data involving human subjects. We demonstrate similar findings with the extended model.

## 5 Soft-max based Flow update for Assorted populations (SoFA)

In this section, we describe our algorithm for solving DAAPs. We provide an algorithm (Algorithm 1) that is general and is applicable to a problem where there are (a) multiple types of agents with each type having a different transition and reward model and (b) multiple notions of rationality. However, for expository purposes (and to focus on the theme of the paper), we consider the case where all agents have the same model (transition and reward functions) and the difference is only in their rationality concept.

---
**Algorithm 1** SoFA($daap, \Gamma, \mathcal{P}$)
---
1: **for all** $\tau \in \Gamma$ **do**
2: $\quad \pi_\tau \leftarrow$ GETINITIALPOLICY($\tau$)
3: **end for**
4: $\tilde{\pi} \leftarrow \phi$
5: $iter \leftarrow 0$
6: **while** $\tilde{\pi} \neq \pi$ **do**
7: $\quad \tilde{\pi} \leftarrow \pi$
8: $\quad$ **for all** $\tau \in \Gamma$ **do**
9: $\quad\quad \langle d^1..d^H \rangle \leftarrow$ AGENTDIST($\pi_{-i}, d^0, \phi$), $i \in \mathcal{P}_\tau$
10: $\quad\quad \{\tilde{x}^t_{s,a}\} \leftarrow$ BESTRSPNSE($\tau, daap, \langle d^1 \cdots d^H \rangle$)
11: $\quad\quad x^t_\tau(s,a) \leftarrow \frac{(iter \cdot x^t_\tau(s,a) + \tilde{x}^t_{s,a})}{iter+1}, \forall s, a, t$
12: $\quad\quad \pi^t_\tau(s,a) \leftarrow \frac{x^t_\tau(s,a)}{\sum_a x^t_\tau(s,a)}, \forall s, a, t$
13: $\quad\quad iter \leftarrow iter + 1$
14: $\quad$ **end for**
15: **end while**
---

Our goal is to compute the policy for perfectly rational agents (policy computed using Algorithm 1) in a population that contains other types of agents. It should be noted that a round-robin best response algorithm does not converge even if we have just one type of perfectly rational agents. Therefore, we introduce our new algorithm, SoFA that is based on the well known Fictitious Play algorithm and has interesting theoretical properties.

Intuitively, the key idea is that at each iteration, an agent computes best response against aggregate policies (over iterations of policy computation) of other agents. Since, in DAAP, the interactions between agents are due to the implicit competition for resources, an agent only has to determine best response to distribution of agent states (and not individual agent states). This observation improves the scalability considerably due to two reasons:
(1) Agents have to plan for state distributions and not against individual states and actions of other agents;
(2) We can assume same policy for all agents of the same type (either due to having same model or same notion of rationality). This allows us to reduce best response computations. Intuitively, it is a reasonable assumption when there is a large number of agents, because such a mixed policy is equivalent to aggregating (differing) individual agent policies.

It should be noted that the best response computation varies from agent to agent depending on their rationality criterion as mentioned in Section 4. Furthermore, best response computation (as mentioned earlier) requires state distribution of other agents. State distributions of other agents can be computed given the starting distribution, type of agents and their policies. Formally, we let (a) $p^t_\tau(s)$ be the probability that an agent of type $\tau$ will be in $s$ at time $t$; (b) $d^t_s$ denote the number of agents in state $s$ at time $t$; (c) $\mathcal{P}_\tau$ denotes the set of players of type $\tau$.

$$p^0_\tau(s) = \frac{d^0_s}{|\mathcal{P}|},$$
$$d^t_s = \sum_{\tau \in \Gamma} p^t_\tau(s) \cdot |\mathcal{P}_\tau|, \forall t \quad (2)$$
$$p^{t+1}_\tau(s) = \sum_{a^t_\tau} \pi^t_\tau(s, a^t_\tau) \sum_{s' \in \mathcal{S}} p^t_\tau(s') \phi^t_{d^t}(s', a^t_\tau, s) \quad (3)$$

For a perfectly rational agent, the best response at each iteration is computed by solving an MDP where the $\phi$ and $\mathcal{R}$ are fixed for the state and action sets because the state distribution of other agents ($d^t, \forall t$) can be computed for each decision epoch using Equation 3. Solving this best response MDP using traditional LP based methods or dynamic programming methods yields a deterministic policy (i.e. one action for each state). Due to this determinism, the aggregation of policies in fictitious play can take many iterations to converge. Therefore, we propose the use of Soft-Max operator for solving MDPs similar to the Soft-Max Value Iteration [15]. This implies that the "max" operator in standard value iteration is replaced by a "soft-max" operator. The soft-max operator makes the strategy equivalent to the quantal response strategy indicated earlier.

We provide the intuition using the following example.

**Example 2** *Consider an MDP with two actions, $a_1$ and $a_2$, each of which can be executed from a state s. $\mathcal{V}(s, a_1)$ and $\mathcal{V}(s, a_2)$ are the expected values for executing actions $a_1$ and $a_2$ respectively from state s. The policy at s with a standard MDP solver (policy iteration, value iteration etc.), $\pi^1$ and SoFA, $\pi^2$ would be:*

$$\pi^1(s, a_1) = \begin{cases} 1, & \text{if } a_1 = \arg\max_{\{a_i \in \{a1, a2\}\}} \mathcal{V}(s, a_i) \\ 0, & \text{otherwise} \end{cases}$$

$$\pi^2(s, a_1) = \frac{e^{\mathcal{V}(s,a_1)}}{e^{\mathcal{V}(s,a_1)} + e^{\mathcal{V}(s,a_2)}}$$

Algorithm 1 provides the pseudo code for the SoFA algorithm. Lines 6-13 provide the core of the algorithm. At each iteration of the algorithm, for each agent there are three key steps: (a) Firstly we compute the state distribution of all other agents (using Equation 3) on line 9; (b) Then, we compute the best response (depends on the agent's rationality criterion) corresponding to the aggregated policies (across iterations) of other agents on line 10; and (c) Finally, we calculate the aggregate state action flows and consequently the aggregate policy for the current agent on lines 11-12.

## 6 Theoretical Results

We provide key theoretical properties of our SoFA algorithm. Firstly, we will show that SoFA converges to equilibrium policies in cases where there are only locally rational agents ($L = 0$) in the mix. Secondly, we show that if SoFA converges on problems with multiple agent types (different transition/reward model or different notions of rationality), then perfectly rational agents will converge to equilibrium policies.

**Proposition 1** *In an assorted population of perfectly rational agents and locally rational agents ($L = 0$), if there exists one type (w.r.t transition and reward model) of perfectly rational agents, then irrespective of the proportion of locally rational agents, SoFA algorithm converges to equilibrium policies for the perfectly rational agents. This is assuming Equation 3 can be used to compute update of probability.*

**Proof.** Let the population, $\mathcal{P}$ be divided into two sets $\mathcal{P}_r$ (perfectly rational agents) and $\mathcal{P}_l$ (locally rational agents), i.e. $\mathcal{P} = \mathcal{P}_l \cup \mathcal{P}_r$. If there exists a potential function for this DAAP, then SoFA would converge to Nash equilibrium. We now define a function $\phi^t()$ and then show that it is a potential function for the DAAP problem. (In the interest of space, we use $\Pi_{ik}$ to represent the set of $\{\pi_i\}_{i \neq k, i \in \mathcal{P}}$ throughout the proof.)

$$\phi^t(\{\pi_i\}_{i \in \mathcal{P}}) = \sum_k \mathcal{V}_i^t(\pi_k, \Pi_{ik}) \quad (4)$$

To show that this is a potential function for a given DAAP, we need to show that for any arbitrary agent $k$ and two of its policies, $\pi_{(k,1)}$ and $\pi_{(k,2)}$:

$$\phi^t(\Pi_{ik} \cup \pi_{(k,1)}) - \phi^t(\Pi_{ik} \cup \pi_{(k,2)})$$
$$= \mathcal{V}_k^t(\pi_{(k,1)}, \Pi_{ik}) - \mathcal{V}_k^t(\pi_{(k,2)}, \Pi_{ik}) \quad (5)$$

We prove this proposition using mathematical induction over the time horizon $t$.

**Base case for t = 0**
For $k \in \mathcal{P}_r$, the value function for an agent is given by:

$$\mathcal{V}_k^0(\pi_{(k,1)}, \Pi_{ik}) = \sum_{s,a} p_k^0(s) \cdot \pi_{(k,1)}(s, a) \cdot \mathcal{R}_k(s, a, d^0)$$

For $k \in \mathcal{P}_l$, the only difference in value function from above would be $d^0 = \mathbf{0}$. It is easy to see that for all agents except $k$ (irrespective of whether $k \in \mathcal{P}_l$ or $k \in \mathcal{P}_r$), the value remains the same if policy does not change for t = 0. Due to this Equation 5 holds (the terms for all other agents will cancel out).

Therefore, let us assume that the $\phi$ is a potential function for horizon t = m, i.e.,

$$\phi^m(\Pi_{ik} \cup \pi_{(k,1)}) - \phi^m(\Pi_{ik} \cup \pi_{(k,2)})$$
$$= \mathcal{V}_k^m(\pi_{(k,1)}, \Pi_{ik}) - \mathcal{V}_k^m(\pi_{(k,2)}, \Pi_{ik}) \quad (6)$$

**Now, we will prove that it holds for $t = m + 1$**

Let us consider the case where $k \in \mathcal{P}_r$,

$$\mathcal{V}_k^{m+1}(\pi_{(k,1)}, \Pi_{ik}) = \sum_{s,a} p_k^0(s) \cdot \pi_{(k,1)}(s, a) \cdot \mathcal{R}_k(s, a, d^0)$$
$$+ \sum_{s,a,s'} p_k^0(s) \cdot \pi_{(k,1)}(s, a) \cdot \phi_k(s, a, s', d^0) \cdot$$
$$\mathcal{V}_k^m(s', \pi_{(k,1)}, \Pi_{ik})$$

From Equation 3

$$= \sum_{s,a} p_k^0(s) \cdot \pi_{(k,1)}(s, a) \cdot \mathcal{R}_k(s, a, d^0) +$$
$$\sum_{s'} p_k^1(s') \cdot \mathcal{V}_k^m(s', \pi_{(k,1)}, \Pi_{ik})$$
$$= \sum_{s,a} p_k^0(s) \cdot \pi_{(k,1)}(s, a) \cdot \mathcal{R}_k(s, a, d^0) + \mathcal{V}_k^m(\pi_{(k,1)}, \Pi_{ik})$$
$$(7)$$

Using Equation 7 and from assumption of Equation 6, we have

$$\mathcal{V}_k^{m+1}(\pi_{(k,1)}, \Pi_{ik}) - \mathcal{V}_k^{m+1}(\pi_{(k,2)}, \Pi_{ik})$$
$$= \sum_{s,a} p_k^0(s) \cdot \pi_{(k,1)}(s,a) \cdot \mathcal{R}_k(s,a,d^0)$$
$$- \sum_{s,a} p_k^0(s) \cdot \pi_{(k,2)}(s,a) \cdot \mathcal{R}_k(s,a,d^0)$$
$$+ \sum_{\{i \neq k, i \in \mathcal{P}\}, s, a} p_i^0(s) \cdot \pi_i(s,a) \cdot \mathcal{R}_i(s,a,d^0)$$
$$- \sum_{\{i \neq k, i \in \mathcal{P}\}, s, a} p_i^0(s) \cdot \pi_i(s,a) \cdot \mathcal{R}_i(s,a,d^0)$$
$$+ \phi^m(\Pi_{ik} \cup \pi_{(k,1)}) - \phi^m(\Pi_{ik} \cup \pi_{(k,2)})$$

Combining terms using the definition of potential function (Equation 4), we have

$$= \phi^{m+1}(\Pi_{ik} \cup \pi_{(k,1)}) - \phi^{m+1}(\Pi_{ik} \cup \pi_{(k,2)})$$

Since $k \in \mathcal{P}_l$ is a sub-case of $k \in \mathcal{P}_r$, where $d^t = \mathbf{0}$, we can adopt the same steps and prove that Equation 5 holds. Hence the proof. ∎

While the following proposition is straightforward, it is important because our algorithm converges on all the problems provided in the experimental results section.

**Proposition 2** *For DAAP problems with multiple types of agents, if SoFA converges, then perfectly rational agents will not have any incentive (in terms of expected value) to deviate.*

**Proof.** If SoFA converges, it implies that the agent policy does not change (line 6) from the previous iteration after computing best response (corresponding to other agent policies). Therefore, perfectly rational agents will have converged to a policy where there is no incentive to deviate in terms of expected value. ∎

## 7 Experimental Results

In this section, we evaluate policies generated by SoFA (perfectly rational if SoFA converges) within the context of the taxi fleet optimization problem, where only a proportion of the taxi drivers follow SoFA strategies and others follow strategies obtained from the cognitive hierarchy model described in Section 4. Firstly, we will perform the behavioral analysis of the real-world taxi data set to obtain the composition of the assorted human-agent population. Once the distributions for assorted human-agent populations are obtained, we will show the performance of SoFA on various synthetic and real-world problem sets.

### 7.1 Human Behavioral Analysis

We now introduce iterative Bayesian inference to compute the distribution over levels of reasoning for taxi drivers. Let $Pr_i^t(L)$ denote the probability of player $i$ using *level-L* reasoning in iteration $t$. Let $f^t(L)$ be the mean probability of *level-L* reasoning in the population, and $f^t(L) = \frac{\sum_i Pr_i^t(L)}{N}$. The expected levels of reasoning after $t$ iterations can be computed by: $mean^t = \sum_L L \cdot f^t(L)$. Let $g_L^t(h)$ denote the perceived proportion of $level - h$ agents by $level - L$ agents in iteration $t$, which can be computed by:

$$g_L^t(h) = \frac{f^t(L)}{\sum_{l=0}^{L-1} f^t(l)}, \forall h < L.$$

Given $\lambda$, $T$, and $\{g_L^t(h)\}$, we can compute the behavioral strategy of *level-L* agent in iteration $t$ by considering actual distribution of taxis. The procedure is based on Bayesian update and its pseudo-code is listed in Algorithm 2. In Algorithm 2, the prior distribution is initialized to be uniform. The movements of free taxis provide necessary observations to continuously update the prior distribution.

The data set we use for computing average levels of reasoning consists of GPS traces from close to 8000 active taxis over 30 working days. Each record in the data set provides a timestamp, taxi coordinate, and the status of the taxi (hired or free). When the taxi is free and moves from one location to another, we consider it as a valid observation. We set the length of time period to be 5 minutes and only consider the morning peak hours from 6 to 9 AM. The data set is sliced into 10 3-day chunks, and the algorithm is executed one chunk at a time. On average, this gives us 90,000 valid observations across all taxis for each run. We also set the look ahead $T = 1$ for behavioral analysis[3].

| $L^{max}$ | $\lambda = 1$ | $\lambda = 3$ |
|---|---|---|
| 4 | $1.21 \pm 0.04$ | $1.50 \pm 0.05$ |
| 5 | $1.35 \pm 0.04$ | $1.58 \pm 0.05$ |

Table 1: Expected levels of reasoning.

Table 1 gives expected value of levels of reasoning for two different but uniform initial distributions and for $\lambda = \{1, 3\}$. $L^{max} = 4$ implies maximum levels of reasoning is set to 4 and each level $\{0 \cdots 4\}$ is assigned initial probability of 0.2. It should be pointed out that although the actual distribution we obtain is not Poisson, the mean values we obtained are well within the

---
[3]As a sanity check for the Bayesian inference, we compared the average revenue of taxis from the data set and using the cognitive hierarchy model. The values were within 1% of each other ($283 and $280 per day respectively).

**Algorithm 2** Iterative Bayesian Inference$(\lambda, T)$

1: $Pr_i^0(L) = \text{UniformDistribution}(L^{\max})$
2: $it \leftarrow 0$
3: **while** $it \leq \text{MaxIterations}$ **do**
4:   $f^{it}(L) = \frac{\sum_i Pr_i^{it}(L)}{N}, \forall L$
5:   $g_L^{it}(h) = \frac{f^{it}(L)}{\sum_{l=0}^{L-1} f^t(l)}, \forall h < L, \forall L$
6:   **for all** DataSet **do**
7:     $\langle d^1 \cdots d^T \rangle \leftarrow \text{GetAgentDist}(\text{DataSet})$
8:     **for all** $L \leq L^{max}$ **do**
9:       $\pi^{it}(L) = \text{Behavior}(\lambda, T, g_L^{it}, \langle d^1 \cdots d^T \rangle)$
10:    **end for**
11:    $s_i = \text{ObservedStartZone}(\text{DataSet})$
12:    $a_i = \text{ObservedAction}(\text{DataSet})$
13:    $Pr_i^{it}(L|a) = \frac{\pi^{it}(L, s_i, a_i) \cdot Pr_i^{it}(L)}{\sum_l \pi^{it}(l, s_i, a_i) \cdot Pr_i^{it}(l)}, \forall L$
14:  **end for**
15:  $it \leftarrow it + 1$
16: **end while**

range of $[1, 2]$ which, as noted by Camerer *et al.*, can explain behavior in close to 100 games. The actual distribution we obtained are skewed towards lower levels of reasoning. A more rigorous study is needed to cross-validate this across larger data set and study the effect of look-ahead, $T$, on the final distribution.

### 7.2 Experimental Setup

To demonstrate that perfectly rational policies are useful in an assorted taxi driver population of different rationalities, we experiment on two different data sets. One is a synthetic data set that simulates random scenarios of demand distributions and another a real-world data set of a taxi company in Singapore. The synthetic data set was created with an inspiration from the real-world taxi data set and so we differentiate between peak hour and normal hours, multiple types of zones, time dependent flow-in and flow-out from a zone etc.

The main parameters of interest in the synthetic data set are (a) the map (defines how zones are connected); (b) total number of zones; (c) number of neighbors for each zone; (d) demand for all zones (represents people going into a zone) characterized by whether it is a residential zone, entertainment zone, office zone or a hospital; (e) total number of taxis; (f) total demand for taxis in a day; (f) length of peak hours. We generate problem instances considering a range of values for all these parameters.

Creating such a detailed simulation data set has helped us in evaluating our algorithm across multiple parameters and investigate the specific parameters affecting the final performance. For each map we compute policies by employing various algorithms and simulate them multiple times to obtain values for the comparison metrics.

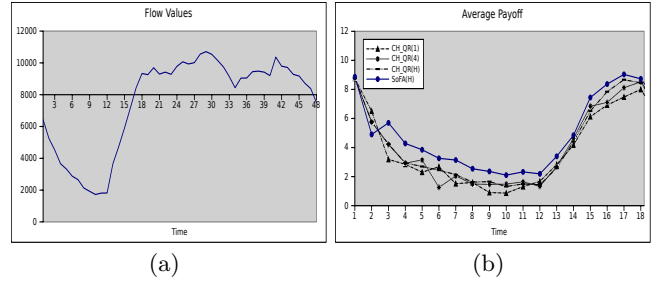

Figure 1: Customer flow at various points and the average payoff obtained by various approaches.

The real-world data set consists of flows of customers, revenues and costs for taxis between various zones across Singapore over a period of six months. We consider a time unit to be 30 minutes long and hence the time horizon for an entire day is $H = 48$. Figure 1 (a) depicts the aggregate demand for taxis throughout the day in this data set. The peak and non-peak hours are easily distinguishable. There are around 8000 taxis and 79 zones.

For all the graphs with aggregated results, we computed standard deviation as well. However, standard deviation is shown in the graphs only if it is greater than 0.1%. We evaluate the algorithms across the social welfare parameters highlighted in Section 2: (1) Average payoff of all taxis. (2) Minimum payoff across all taxis and (3) Starvation, which represents the lack of availability of taxis. The first two metrics provide an intuition for how the payoff distribution for taxi drivers changes. The third metric ensures an improvement from the system management (or taxi fleet owners) perspective.

We first present the results for the case when all agents adopt one type of rationality and then we will present the results for an assorted population. Although, we do not have theoretical guarantees on convergence when there are agents following behavioral strategies along with perfectly rational agents, empirically, we achieved convergence on all the problem instances. Our algorithm converged on all the problems in the synthetic and real-world data set within 2 hours[4].

### 7.3 Homogeneous Population

We compare the performance of SoFA algorithm against the strategies generated by the extended cognitive hierarchy model described in Section 7.1. While we experimented with a varied set of $\lambda$ values, we only show results for $\lambda = \{1, 3\}$ because they yield the most diverse set of observations. As for the time horizon, we considered $T = \{1, 4, H\}$. As for the levels of reasoning, $L$, we use the distributions obtained from iterative Bayesian inference corresponding to $\lambda = \{1, 3\}$. We also set $L^{max} = 4$. Thus, in our result graphs, $CH\_QR(5)$ with $\lambda = 3$ would imply the population consists of $\{17\%, 39\%, 26\%, 13\%, 5\%\}$ of levels $\{0 \cdots 4\}$ agents respectively with *look-ahead=5*. For $\lambda = 1$, it would be $\{20\%, 52\%, 17\%, 5\%, 6\%\}$.

---
[4]The most amount of time was taken by the real-world problem set, where there are 8000 taxis and 79 zones.

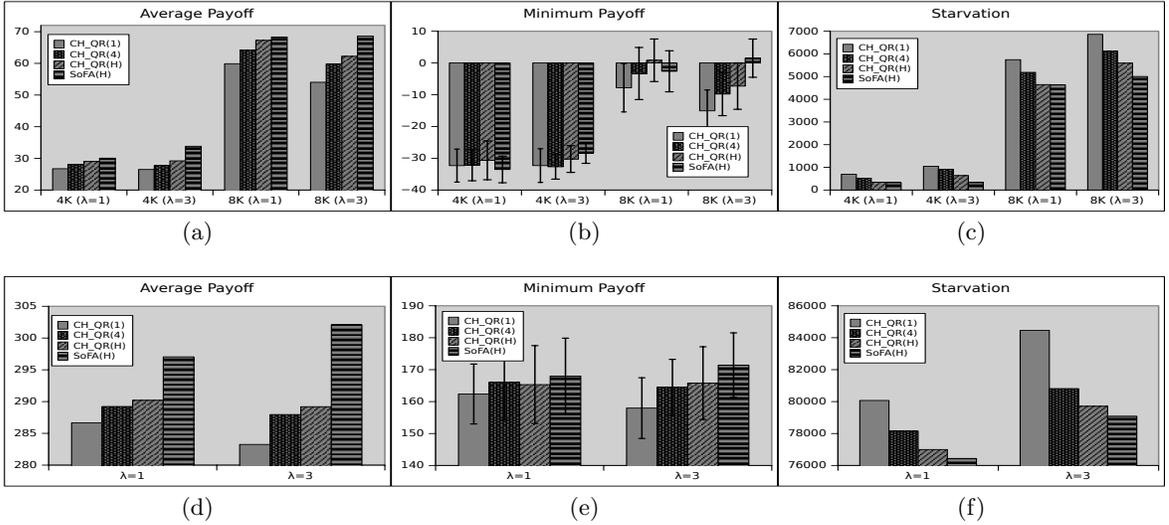

Figure 2: Performance of SoFA and behavioral strategies with $T = \{1, 4, H\}$ on synthetic and real-world datasets.

Figure 2 provides the results on the synthetic and real-world data sets. Graphs (a)-(c) are results for 40 zone problems and (d)-(f) are for the the taxi data set. Here are some of the key observations with respect to average and minimum payoff, starvation:

(i) For both synthetic and real-world problems, SoFA outperforms all the benchmark algorithms, due to the Nash equilibrium policies employed by SoFA agents.
(ii) Behavioral strategies with look ahead equal to horizon outperformed strategies with shorter look ahead.
(iii) We observe that for behavioral strategies, $\lambda = 1$ performs better than $\lambda = 3$. This could be because for $\lambda = 3$, the proportion of higher level reasoning agents increases and hence overall performance goes down. On the other hand, for SoFA, due to the randomized policies, performance increases from $\lambda = 1$ to $\lambda = 3$.
(iv) For the synthetic data set, when population size is $4k$, the behavioral strategy with $\lambda = 1$ and look ahead equals to *horizon* obtains marginally higher minimum payoff than all other algorithms. But when the population size is increased to $8k$ and with $\lambda = 3$, minimum payoff is guaranteed to be orders of magnitude higher under SoFA.
(v) For synthetic and real-world data sets, SoFA outperforms others for $\lambda = 3$ w.r.t. minimum payoff.
(vi) With respect to starvation (lower values are better), SoFA outperforms others on all data sets.

Figure 1(b) gives the average payoff for each algorithm at each time step on the real-world data set. We observe that SoFA performs better than other algorithms. Furthermore, it should be noted that there is high degree of similarity in the performance of SoFA over time and the flow values of Figure 1(a). This implies that SoFA is able to adapt quickly to the changes in demand over time.

### 7.4 Assorted Population

We now evaluate SoFA with an assorted population of agents. It should be noted that for a fixed $\lambda$ and $T$, there are more than one type of agents in the population. We use X to denote the ratio of agents that are following SoFA policies. For instance, X = 20% (along X-axis in Figure 3) implies 20% agents are perfectly rational and the rest are following behavioral strategies for a particular $\lambda$.

Figure 3 provides the performance of SoFA within a varying population of agents: (a)-(c) and (d)-(f) are results for the synthetic and real-world data sets respectively. We make the following observations regarding average payoff, minimum payoff, and starvation:

(i) The performance of SoFA improves as the population of agents following behavioral strategies increases from 20% to 80%. The performance of boundedly rational agents at the same time goes down. It indicates that irrespective of the population of boundedly rational agents, it is always useful to employ a SoFA policy. SoFA outperforms behavioral strategies even when the proportion of agents following SoFA increases to 80%.
(ii) On the synthetic data set, SoFA outperforms other algorithms w.r.t. minimum revenue when the propor-

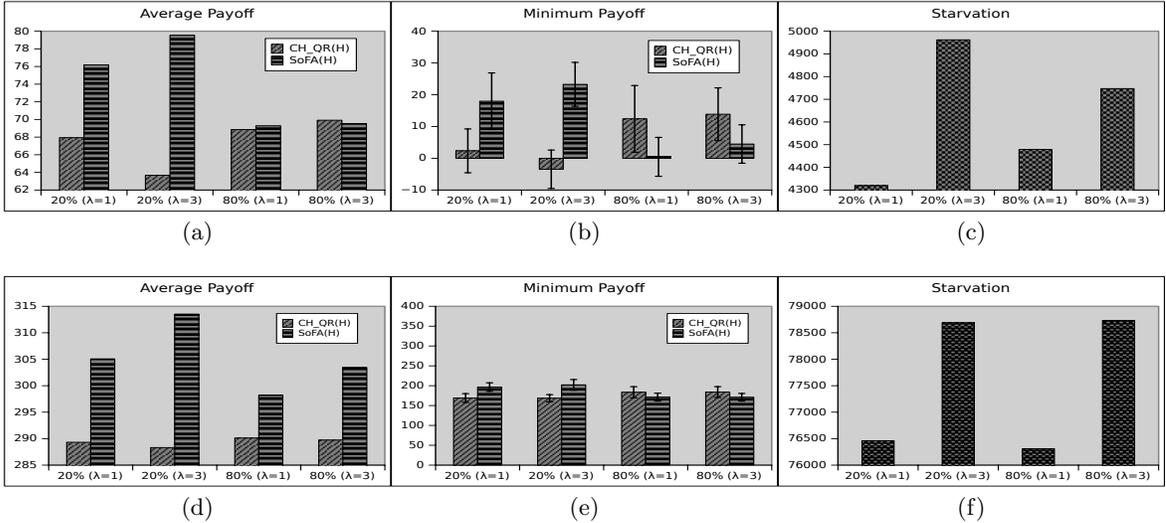

Figure 3: SoFA vs behavioral strategies in assorted population with $T = H$ on synthetic and real-world datasets.

tion of agents is 20%.
(iii) On real-world data set the minimum payoffs are almost equal across both strategy models.
(iv) Starvation values decrease with increase in the number of perfectly rational agents.

From all these experiments on both synthetic and real-world problem instances, we conclude that with respect to all welfare metrics (average revenue, minimum revenue and starvation), for any agent, it is useful to pursue the policy computed for perfectly rational agents in problems such as taxi fleet optimization.

## 8 Related Work and Conclusions

In this section, we briefly describe research related to the contributions made in this paper. The first thread of related research is in the field of transportation. User equilibrium (UE) is a classical and powerful equilibrium concept in transportation explaining individual route choices in face of competition for road usages from other users. Originally proposed in static setting [10], it was later expanded to dynamic cases (where temporal choices are also important) [3]. In either format, static or dynamic, the concept of UE provides a way to infer and to predict the behaviors of individual drivers; such ability helps not just individual drivers to identify better routes, but it can also help policy makers to properly design road network in anticipation of driver's responses. While the solution concept of UE is relevant, it does not account for the presence of involuntary movements for agents.

The second thread of related research is from *behavioral game theory* where different models have been proposed to accommodate experimental results (involving human subjects) and theoretical predictions. These include cognitive hierarchy model [1] and quantal response equilibrium [4]. More recently, variants of these models have been evaluated across different games [13]. As opposed to evaluation on traditional small benchmark problems, we have studied these models in the context of a very large scale real-world setting in this paper.

The next thread of relevant research is due to Weintraub *et al.* [12, 11]. This research introduces the concept of oblivious equilibrium for large scale dynamic games. They provide a mean field approximation to solve problems where there is stochasticity in state transitions. While, the problem is similar to DAAPs, the assumption of mean field (or a stationary distribution of taxis in our case) is not applicable in the context of taxi problems. In fact, there is a huge variance in the set of possible distributions at each decision epoch and hence oblivious equilibrium is not directly applicable in our context.

DAAP model represents a subset of problems represented by the generic Partially Observable Stochastic Games (POSG) model. However, all the approaches [6, 9, 8] provided in the literature are for solving identical payoff stochastic games (also referred to as Decentralized POMDPs or DEC-POMDPs) and not generic POSGs. Furthermore, they do not scale to problems with 8000 agents.

### Acknowledgement

The research described in this paper was funded in part by the Singapore National Research Foundation (NRF) through the Singapore-MIT Alliance for Research and Technology (SMART) Center for Future Mobility (FM).

# References

[1] C. F. Camerer, T.-H. Ho, and J.-K. Chong. A cognitive hierarchy model of games. *Quarterly Journal of Economics*, 119(3):861–898, 2004.

[2] J. Filar and K. Vrieze. *Competitive Markov decision processes*. Springer-Verlag New York, Inc., New York, NY, USA, 1996.

[3] T. L. Friesz, D. Bernstein, T. E. Smith, R. L. Tobin, and B. W. Wie. A variational inequality formulation of the dynamic network user equilibrium problem. *Operations Research*, 41(1):179–191, 1993.

[4] R. D. McKelvey and T. R. Palfrey. Quantal response equilibria for normal form games. *Games and Economic Behavior*, 2:6–38, 1995.

[5] R. W. Rosenthal. A class of games possessing pure-strategy Nash equilibria. *International Journal of Game Theory*, 2(1):65–67, 1973.

[6] S. Seuken and S. Zilberstein. Improved memory-bounded dynamic programming for decentralized POMDPs. In *UAI*, 2007.

[7] P. Varakantham, S.-F. Cheng, and T. D. Nguyen. Decentralized decision support for an agent population in dynamic and uncertain domains. In *10th International Conference on Autonomous Agents and Multiagent Systems*, pages 1147–1148, 2011.

[8] P. Varakantham, J. Y. Kwak, M. Taylor, J. Marecki, P. Scerri, and M. Tambe. Exploiting coordination locales in distributed POMDPs via social model shaping. In *ICAPS*, 2009.

[9] P. Velagapudi, P. Varakantham, K. Sycara, and P. Scerri. Distributed model shaping for scaling to decentralized POMDPs with hundreds of agents. In *10th International Conference on Autonomous Agents and Multiagent Systems*, pages 955–962, 2011.

[10] J. G. Wardrop. Some theoretical aspects of road traffic research. In *Proceedings of the Institute of Civil Engineers - Part II*, volume 1, pages 325–378, 1952.

[11] G. Weintraub, C. Benkard, and B. V. Roy. Markov perfect industry dynamics with many firms. *Econometrica*, 76(6):1375–1411, 2008.

[12] G. Y. Weintraub, L. Benkard, and B. V. Roy. Oblivious equilibrium: A mean field approximation for large-scale dynamic games. In *NIPS*, 2006.

[13] J. R. Wright and K. Leyton-Brown. Beyond equilibrium: Predicting human behavior in normal-form games. In *Twenty-Fourth Conference of the Association for the Advancement of Artificial Intelligence (AAAI-10)*, pages 901–907, 2010.

[14] H. Yang and S. C. Wong. A network model of urban taxi services. *Transportation Research Part B: Methodological*, 32(4):235–246, 1998.

[15] B. D. Ziebart. *Modeling Purposeful Adaptive Behavior with the Principle of Maximum Causal Entropy*. PhD thesis, Carnegie Mellon University, 2010.